\documentclass[conference,10pt,letter]{IEEEtran}

\usepackage{cite}

\usepackage{amsmath,amsthm}
\usepackage{amssymb, bm}
\usepackage{float}
\usepackage{amssymb,amsmath,amsfonts}
\usepackage{graphicx}
\usepackage{color}

\newcommand{\modop}[2]{\left(\left({#1}\right)\right)_{#2}}
 
\newcommand{\e}{\varepsilon}
\newcommand{\ua}{\underline{\alpha}}
\newcommand{\ub}{\underline{\beta}}
\newcommand{\BP}{\textsc{\tiny BP}}
\newcommand{\MAP}{\textsc{\tiny MAP}}
\newcommand{\opt}{\textsc{\tiny opt}}

\begin{document}
\title{Triggering Wave-Like Convergence of\\ Tail-biting Spatially Coupled LDPC Codes}

\author{\IEEEauthorblockN{Sebastian Cammerer\IEEEauthorrefmark{1}, Vahid Aref\IEEEauthorrefmark{2}, Laurent Schmalen\IEEEauthorrefmark{2}, and Stephan ten Brink\IEEEauthorrefmark{1} }
\IEEEauthorblockA{
\IEEEauthorrefmark{1} Institute of Telecommunications, University of  Stuttgart, Stuttgart, Germany 
\\
\IEEEauthorrefmark{2} Bell Laboratories, Alcatel-Lucent, Stuttgart, Germany
}
}

\maketitle

\begin{abstract}
Spatially coupled low-density parity-check (SC-LDPC) codes can achieve the channel capacity under low-complexity belief propagation (BP) decoding, however, there is a non-negligible rate-loss because of termination effects
for practical finite coupling lengths. In this paper, we study how we can approach the performance of terminated SC-LDPC codes by random shortening of tail-biting SC-LDPC codes. We find the minimum required rate-loss in order to achieve the same performance than terminated codes. We additionally study the use of tail-biting SC-LDPC codes for transmission over parallel channels (e.g., bit-interleaved-coded-modulation (BICM)) and investigate how the distribution of the coded bits between two parallel channels can change the performance of the code. We show that a tail-biting SC-LDPC code can be used with BP decoding almost anywhere within the achievable region of MAP decoding. The optimization comes with a mandatory buffer at the encoder side. We evaluate different distributions of coded bits in order to reduce this buffer length.
\end{abstract}

\IEEEpeerreviewmaketitle

\section{Introduction}

Their excellent performance under low-complexity belief propagation (BP) decoding 
renders spatially coupled LDPC codes (SC-LDPC) attractive for error correction subsystems in upcoming communication systems such as, e.g., long-haul optical fiber transceivers~\cite{schmalen2015spatially}.
It was proven in \cite{Coupl11bec,Coupl11BMS,yedla2012simple,kumar2012proof} that SC-LDPC codes
can achieve the capacity of binary-input, memoryless, symmetric-output (BMS) channels under BP decoding. In particular, it was proven that the BP threshold of an SC-LDPC ensemble asymptotically converges to the threshold of the underlying LDPC ensemble under maximum-a-posteriori (MAP) decoding. This phenomenon is called \emph{threshold saturation}. 

The idea of spatial coupling is to replicate $L\gg 1$ copies of an LDPC code along a \emph{spatial dimension} and connecting the individual LDPC codes by \emph{swapping edges} in a way that the local graphical structure of the original code is preserved. Terminating the replicas in an effective way triggers a ``decoding wave''~\cite{aref2013convergence} and threshold saturation occurs. 
Termination is usually performed by setting \emph{all} code bits of a few LDPC copies at the boundaries of the spatial chain to zero~\cite{lentmaier2010iterative,Coupl11BMS}, reducing the total encoding rate. Although this rate-loss vanishes as $L\to\infty$, it is not negligible for  practical finite values of $L$.

Using tail-biting SC-LDPC codes~\cite{tavares2007tail} is a promising way to mitigate the rate-loss for a practical finite $L$. Threshold saturation occurs if there are enough reliably received code bits in just a few spatial positions. First, these reliable code bits are decoded and then, the decoder progresses successively along the spatial dimension. These  reliable code bits can be obtained by shortening, as suggested (however without analysis) in~\cite{Coupl10BMS}.

If transmission takes place over different parallel channels, the required reliable code bits can be provided without shortening (and rate-loss) by carefully interleaving the code bits among channels with different entropies. Examples of parallel channels include bit-interleaved coded-modulation (BICM) for high spectral efficiency modulation formats and multi-carrier transmission, such as orthogonal frequency division multiplexing (OFDM) or transmission over two polarizations, frequently occurring in optical communications. This approach has been first proposed in~\cite{hager2014optimized,hager2015terminated} where optimized bit interleavers for BICM have been numerically derived based on a Gaussian approximation of log-likelihood ratio (LLR) densities of channels and a P-EXIT approximation~\cite{liva2007protograph} for density evolution.

In this paper, we study both above approaches in a unified framework. 
We also investigate the effect of an additional practical constraint, 
the encoding buffer size required for interleaving. 
We use the technique of density evolution (DE) over a BEC to minimize the total amount of shortening. We show that the results almost immediately carry over to a binary additive white Gaussian noise (BAWGN) channel. In a second part, we consider the scenario of parallel channels. Using DE, we numerically find the 
achievable region of two BECs under BP decoding.
As a more practical case, we revisit the practical application of BICM over the AWGN channel. In comparison to \cite{hager2014optimized,hager2015terminated}, 
we apply a more precise numerical method for DE~\cite[App. B]{URbMCT} and no Gaussian approximation of channel densities. 
One noteworthy observation is that the tail-biting SC-LDPC code with an optimized bit-interleaving outperforms the terminated SC-LDPC codes with the same parameters and the same energy per bit. However, it is not clear if this generally applies for all tail-biting SC-LDPC codes.
Finally, we show  how the performance is degraded and how the optimal 
interleaving needs to be modified if accounting for a size-constrained encoding buffer.

\section{Tail-biting Spatially Coupled LDPC Codes}
\label{sec:tail-biting_sc_ldpc}
To construct tail-biting SC-LDPC codes, we first lay out a set of \emph{spatial} positions indexed 
by integers $z\in[0,L)$ on a circle, where $L$ denotes the \emph{replication factor}. 
We fix a ``smoothing parameter'' $w \in \mathbb{N}$. To each position $z$, we assign $n$ code bits and $m$ parity check constraints (hence in total $N=Ln$ code bits and $M=Lm$ parity constraints). 
We assign a variable node to each code bit and 
assign a check node to each parity constraint. 

To construct a random instance of the tail-biting SC-LDPC$(d_v,d_c,L,w,n)$ ensemble, 
we connect the variable nodes and the check nodes
in the following manner: 
Each variable node at position $z$ is connected randomly to $d_v$ check nodes lying within the range $\modop{[z,z+w-1]}{L}$, where $\modop{x}{L}$ returns the remainder of the integer division of $x$ by $L$ (``$x$ modulo $L$'').
Equivalently, each check node at position $z$ is connected
randomly to $d_c$ variable nodes in the range $\modop{[z-w+1,z]}{L}$.
For additional details, we refer the reader to \cite{Coupl11BMS,Arefthesis}. If all code bits in positions $z\in[0,w-2]$ are fixed to be zero, the code ensemble becomes a (terminated) SC-LDPC ensemble~\cite{Coupl11BMS}.

When $n\to\infty$, it is common to use 
 density evolution (DE)~\cite{URbMCT} 
to evaluate the asymptotic performance of the code ensemble
under BP decoding. 
This method estimates how the empirical
distribution of the LLRs of code bits
 at each position $z$ evolves iteratively during BP decoding, given
the empirical distribution of the received bits' LLRs.
The DE equations for SC-LDPC codes over a BMS channel are detailed in \cite{Coupl11BMS}.
For the case of BICM and BAWGN channels,
we use the numerical DE method in \cite[App.~B]{URbMCT}. 
For transmission over BECs, the LLR distribution can be represented by a scalar value, the erasure probability, and 
the DE equation turns into a scalar update recursion~\cite{Coupl11bec}.
In that case, the update equation becomes
\begin{equation*}
x^{(t+1)}_z = \e_z \left(\!1\!-\!\frac{1}{w}\sum_{i=0}^{w-1}\left(1\!-\!\frac{1}{w}\sum_{j=0}^{w-1} x^{(t)}_{\modop{z+i-j}{L}}\!\!\right)^{d_c-1} \right)^{d_v-1}
\end{equation*}
where $\e_z$ denotes the average erasure probability of code bits at spatial position $z$ and $x^{(t)}_z$ denotes the average erasure probability of the outgoing messages from code bits in position $z$ and at iteration $t$. We initialize $x_z^{(0)}=1$ for all $z\in[0,L)$. For details, see~\cite{Coupl11bec}.

\section{Optimal Shortening Over a BMS}
\label{sec:random_shortening}

\begin{figure}[tb]
\centering
\includegraphics[width=0.7\columnwidth]{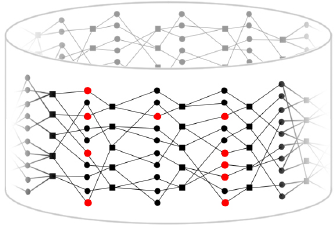}
\caption{\label{fig:non_uniform_alpha}Tanner graph of a tail-biting SC-LDPC$(d_v=2,d_c=4,L=12,w=2,n=10)$ code. Red variable nodes show shortened code bits.
 }
\end{figure}

Consider the tail-biting SC-LDPC$(d_v,d_c,L,w)$ ensemble.
It is known that the BP threshold and the MAP threshold of this ensemble are equal to the BP threshold
and the MAP threshold of the underlying 
LDPC$(d_v,d_c)$ ensemble~\cite{Coupl11BMS}. With a 
properly selected shortening, the BP threshold of 
the spatially coupled ensemble saturates to the MAP threshold. We shorten the code by setting 
some code bits to zero, as indicated in Fig.~\ref{fig:non_uniform_alpha} by red variable nodes. We thus inject some prior knowledge that can be used to
trigger BP decoding.
Let $\alpha_z$ be the fraction of \emph{randomly} shortened code bits (i.e., they are fixed to zero) for  $z\in[0,L)$. 
For example, we recover the terminated SC-LDPC$(d_v,d_c,L,w)$ ensemble for $\alpha_z=1$ for $0\leq z\leq w-2$ and $\alpha_z=0$ otherwise.
 This random shortening decreases the design rate of the code from $R=1-d_v/d_c$ to
\begin{equation}
\label{eq:rate}
R(\ua) = 1 - \frac{d_v}{d_c}\cdot\frac{L-\sum_{z=0}^{L-1}
\left(\frac{1}{w}\sum_{j=0}^{w-1}\alpha_{(({z-j}))_L}\right)^{d_c}}{L-\sum_{z=0}^{L-1}\alpha_z},
\end{equation}
where $\ua=(\alpha_0,\dots,\alpha_{L-1})$. 
For simplicity,
we explain our optimization method for transmission over a BEC
with erasure probability~$\e$. In this case,
the average fraction of code bits erased  by the channel
at position $z$ is $\e_z=(1-\alpha_z)\e$. Equivalently,
the code bits at position $z$ "see" a BEC with erasure probability~$\e_z$. Let $P_e(\ua,T)$ denote the bit erasure probability of BP decoding after $T$ iterations.

Let $\e_{\BP}$ and $\e_{\MAP}$ denote the BP threshold and the MAP threshold of 
the uncoupled LDPC$(d_v,d_c)$ ensemble. 
For any $\e\in(\e_{\BP},\e_{\MAP})$, 
our goal is to find $\ua$ which maximizes $R(\ua)$ under successful BP decoding, i.e.,
\begin{equation}
\ua_{\opt}(\e) = \arg\max_{\ua}\{R(\ua)\mid \lim_{T\to\infty} P_{\rm e}(\ua,T)=0\}.
\end{equation}
For numerical computation, we relax this maximization:
\begin{itemize}
\item[(i)] For a given $\ua$, define $T_{\delta}(\ua)$ as the smallest iteration $t$ such that $\frac{1}{L}\sum_{z=0}^{L-1} \vert x^{(t-1)}_z-x^{(t)}_z\vert<\delta$.
\item[(ii)] We find
\end{itemize}
\begin{equation}
\label{eq:relaxed_Pe}
\ua^*(\e) = \arg\max_{\ua}\{R(\ua)\mid P_{\rm e}(\ua,T_\delta(\ua)) <\delta\}.
\end{equation}

\begin{figure}[tb]
\includegraphics[width=\columnwidth]{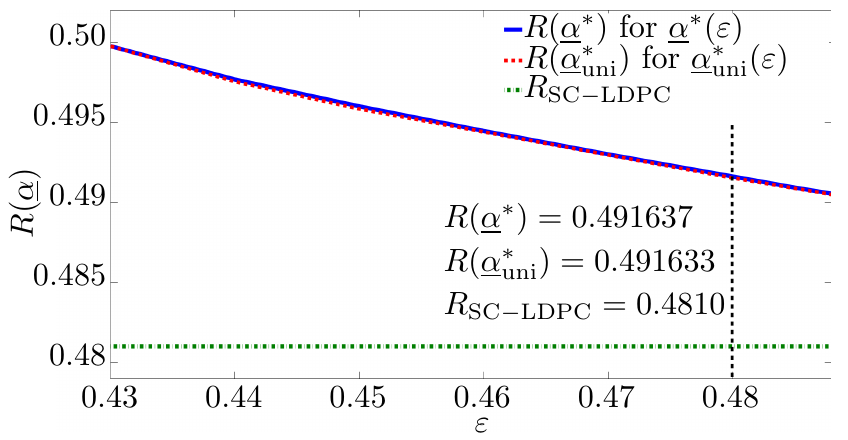}
\caption{\label{fig:R_alpha} The design rate $R(\underline{\alpha})$ of the
tail-biting SC-LDPC$(3,6,50,3)$ ensemble 
shortened
according to either $(i)$ the optimized $\underline{\alpha}^*(\e)$
(blue curve) or $(ii)$
$\underline{\alpha}_{\rm uni}^*(\e)$ (red curve).
These curves are compared with the design rate of
the terminated SC-LDPC$(3,6,50,3)$ ensemble (green curve).}
\end{figure}

We fix $\delta=10^{-7}$ throughout the paper.
To find $\ua^*(\e)$, we apply two sub-optimal algorithms: 
(i) For small $L$, an exhaustive search is carried out over the discretized space of ${\ua}$. Each $\alpha_z$ is quantized with resolution $\Delta=10^{-3}$. (ii) The sub-optimal differential evolution algorithm~\cite{storn1995differential} is also used for large $L$ which is generally much faster than an exhaustive search. We observe a good consistency between the results of both algorithms. For example, 
the optimized $\ua^*=(0.324,0.149,0.350,0,\dots,0)$ is obtained for the tail-biting SC-LDPC$(d_v=3,d_c=6,L=50,w=3)$ ensemble used over a BEC with $\e=0.48$.
We also consider a simpler scheme denoted ``uniform shortening'', defined as
\begin{equation}
\ua_{\rm uni}(B) =
\begin{cases}
\alpha_z=\alpha,& 0\le z< B\\
\alpha_z=0,& \text{otherwise}
\end{cases}
\label{eq:uniform_alpha}
\end{equation}
where $0\leq B < L$. Similarly, we can find  
\begin{equation}
\label{eq:uniform_opt}
\ua_{\rm uni}^*(\e)=\arg\max_{B,\ua_{\rm uni}(B)}\{R(\ua)\mid P_{\rm e}(\ua,T_\delta(\ua))<\delta\}.
\end{equation}
The above optimization is much simpler. For each $B$, it is a one-dimensional optimization over a bounded interval. The BP performance is monotonic in $\alpha$, and thus,
we can simply use algorithms such as the bisection method to find the best $\ua_{\rm uni}(B)$. Then we change $B$ to find $\ua_{\rm uni}^*(\e)$.
Fig.~\ref{fig:R_alpha} illustrates the optimization results for the tail-biting SC-LDPC$(3,6,50,3)$ ensemble. The maximum design rate is computed for both $\ua^*(\e)$
and $\ua_{\rm uni}^*(\e)$, and for $\e\in(\e_\BP,\e_\MAP)$. We observe that both optimized shortenings
reduce the rate-loss by more than $50\%$. Moreover, there is only a very small difference 
between the rate of $\ua^*(\e)$ and $\ua_{\rm uni}^*(\e)$.
We observe the same behavior for tail-biting ensembles of LDPC$(4,8)$ and LDPC$(5,10)$ codes. 
Note that the error probability of BP decoding is monotonically increasing in terms of $\e$. Therefore, $\ua^*(\e_\MAP)$ is also feasible for $\e\le \e_\MAP$. 
This $\ua^*(\e_\MAP)$ can be universally used for all $\e\in(\e_\BP,\e_\MAP)$ with more than $50\%$ rate-loss gain, even though it is not the best $\ua$ for~$\e<\e_\MAP$.

\begin{figure}[tb]
\centering
\includegraphics[width=\columnwidth]{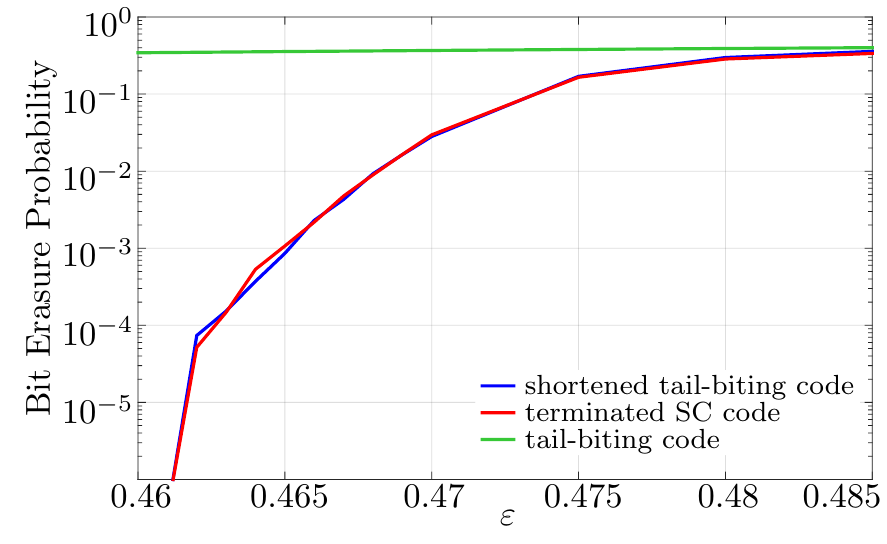}
\caption{
\label{fig:ber_bec} BP performance in terms of channel erasure probability $\e$ for a shortened tail-biting SC-LDPC code (blue curve), a terminated SC-LDPC code (red curve), and a tail-biting SC-LDPC code (green curve). For all codes, 
$(d_v,d_c,L,w,n)=(3,6,50,3,2000)$. }

\end{figure}

\begin{figure}[tb]
\centering
\includegraphics[width=\columnwidth]{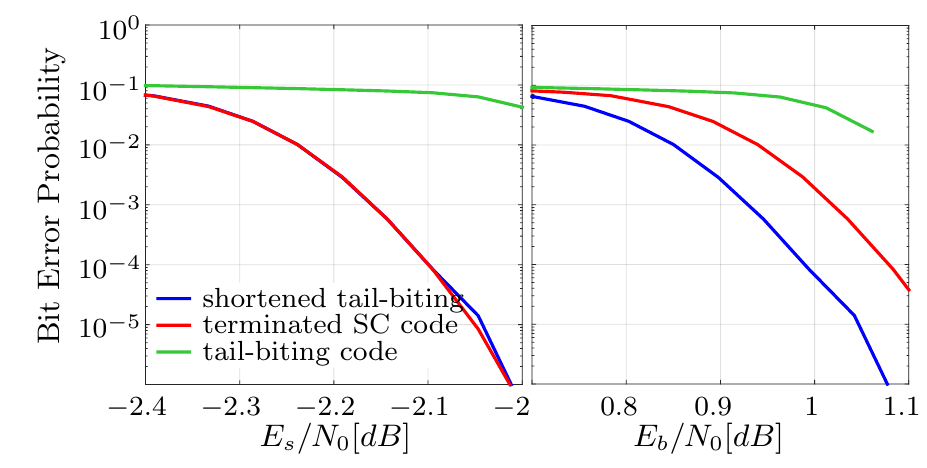}
\caption{\label{fig:ber_awgn} BP performance in the  BAWGN channel for a shortened tail-biting SC-LDPC code (blue curve), a terminated SC-LDPC code (red curve), and a tail-biting SC-LDPC code (green curve). For All codes, $(d_v,d_c,L,w,n)=(3,6,50,3,2000)$. Left: BER over $E_s/N_0$. Right:
BER over $E_b/N_0.$}
\end{figure}

Fig.~\ref{fig:ber_bec} shows the average bit erasure probability $P_e$ under BP decoding for
the tail-biting SC-LDPC$(3,6,50,3)$ code shortened by $\ua^*(\e_\MAP)$. For comparison, we also plot
the erasure probability curve of the terminated SC-LDPC$(3,6,50,3)$ code, and the tail-biting code
(without shortening). For all codes, $n=2000$. 
We observe that the terminated SC-LDPC code has much smaller $P_e$ than the tail-biting SC-LDPC code as 
the BP threshold of SC-LDPC code is increased to the MAP threshold. With shortening, 
not only the performance of the tail-biting SC-LDPC code improves to the performance of the terminated SC-LDPC code but also the shortened tail-biting SC-LDPC code has a larger rate than the SC-LDPC code.
We also plot the performance of these codes over a binary additive white Gaussian noise (BAWGN) channel
in Fig.~\ref{fig:ber_awgn}. The left sub-plot shows the bit error rate (BER) in terms of 
signal-to-noise (SNR) ratios $E_s/N_0$. We observe the similar behavior as for BEC in Fig.~\ref{fig:ber_bec}. In order to see the gain in coding rate, 
the same BER values are plotted in terms of $E_b/N_0=E_s/(RN_0)$ in the right sub-plot. It shows that 
the larger rate leads to an additional \emph{net coding gain} of $\approx 0.1$\,dB.

\section{Bit Interleaving Over Two Parallel Channels}
In this section, we consider the transmission over two independent parallel channels. We show that carefully 
interleaving code bits of tail-biting SC-LDPC code between channels leads to significantly improved BP decoding performance without rate-loss. The results can be extended to more than two parallel channels in a straightforward manner.

\subsection{Parallel BECs}\label{sec:two_becs}

We explain our optimization method
using two independent parallel BECs with erasure probabilities $\e_1$ and $\e_2$ and where both channels shall be used an equal number of times.
In this case, the low complexity of DE analysis allows us to 
evaluate the complete achievable region of $(\e_1,\e_2)$ under BP decoding and find the optimal bit interleaving.

We assume that the transmitter knows the channel parameters $\e_1$ and $\e_2$ and w.l.o.g. that $\e_1\le\e_2$.
The code bits are transmitted over either the first or the second channel. 
If exactly half of the code bits in each spatial position are passed through each channel, this is equivalent
to transmitting over a BEC with erasure probability $\frac{1}{2}(\e_1+\e_2)$.  In this case, without termination,
the tail-biting SC-LDPC ensemble cannot be decoded successfully unless $\frac{1}{2}(\e_1+\e_2) < \e_\BP$. 

\begin{figure}[tb]
\centering
\includegraphics[width=0.7\columnwidth]{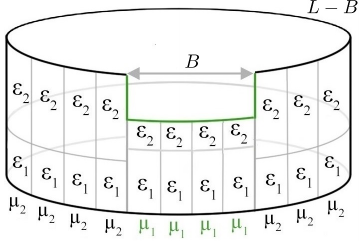}
\caption{Interleaving the code bits of different spatial positions between two parallel independent BECs such that for the average it holds $\mu_1<\mu_2$.}
\label{separation_epstop_epsbottom}
\end{figure}

We show that we can exceed this bound if we carefully interleave the code bits of different spatial positions between channels.
Let $\beta_{z}$  denote the fraction of code bits in spatial position $z$ 
transmitted over the BEC with $\e_1$. Clearly, $\sum_{z=0}^{L-1} \beta_{z}=L/2$.
The average erasure probability at position $z$ is $\e_z=\beta_z \e_1 + (1-\beta_z)\e_2$. Now, the problem of finding optimal 
$\beta_z$ becomes similar to the best shortening optimization presented
in Section~\ref{sec:random_shortening}.
We already observed in Fig.~\ref{fig:R_alpha} that there is a very small difference between 
uniform shortening and non-uniform shortening. 
Due to its simplicity, we employ an approach similar to uniform shortening and consider the setup illustrated in Fig.~\ref{separation_epstop_epsbottom}:
\begin{equation*}
\ub_{\rm uni}(B) =
\begin{cases}
\beta_0,&\hspace*{-.4cm} 0\le z< B\\
\frac{\frac{L}{2}-B\beta_0}{L-B},&\hspace*{-.1cm} z\ge B
\end{cases}\Rightarrow
\e_z=
\begin{cases}
\mu_1,& 0 \leq z< B\\
\mu_2,& z\geq B
\end{cases}
\end{equation*}
for some $B<L/2$.
It becomes now equivalent to the shortened code over a single BEC with erasure probability $\mu_2$ with uniform shortening $\alpha=\mu_1/\mu_2$ in Eq.~\eqref{eq:uniform_alpha} .
It is easy to check $\mu_2\leq \frac{(\e_1+\e_2)/2-(B/L)\e_1 }{(1-B/L)}$ and $\e_1\leq\mu_1$. If we know all the
feasible 3-tuples $(\mu_1,\mu_2,B)$, we can determine the achievable region $(\e_1,\e_2)$.
The idea is as follows: if $\mu_1<\e_\BP$ and $B$ is large enough, 
the BP algorithm can decode some code bits lying in the range $z<B$. In particular,
the ``computation graph'' of some code bits has enough non-erased variable nodes
for the BP algorithm to recover them. Those recovered code bits then play the role of the effective termination
for the coupled ensemble if $\mu_2$ is small enough (i.e.,  $\mu_2<\e_\MAP$). Note that
successful BP decoding is not attained by merely interleaving
if either $\e_2\ge\e_1\ge\e_\BP$ or $\e_1+\e_2\ge 2\e_\MAP$.

For a given $(\mu_2,B)$, let $f(\mu_2,B)$ denote the largest $\mu_1$ for which
BP decoding succeeds (see \eqref{eq:uniform_opt}). We have 
$\e_1\leq\e_\BP$ and
\begin{align}
2\left(1-\frac{B}{L}\right)\mu_2 &\leq \left(1-\frac{2B}{L}\right)\e_1 + \e_2 \nonumber\\
\e_1 + \e_2 &\leq \frac{2B}{L} f(\mu_2,B) + 2\left(1-\frac{B}{L}\right)\mu_2
\label{eq:g_(B,mu)}
\end{align}

\begin{figure}[tb]
\centering
\includegraphics[width=0.8\columnwidth]{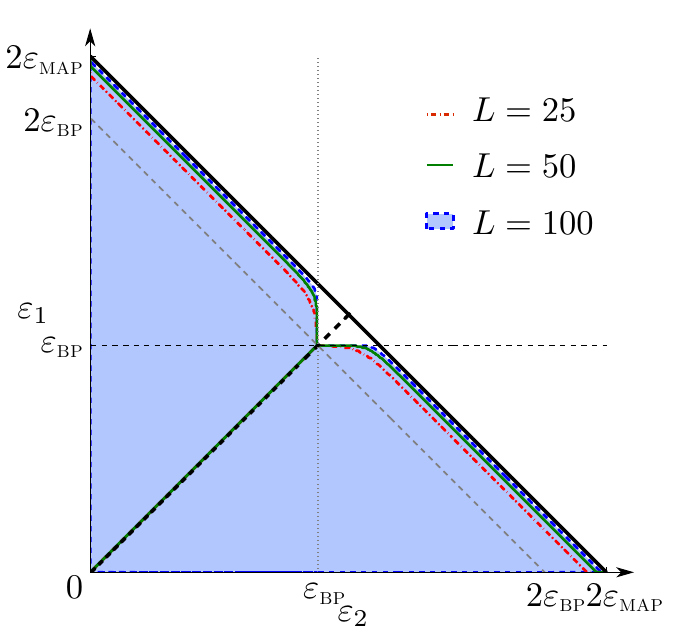}
\caption{\label{fig:areas} The achievable region under BP decoding for the tail-biting SC-LDPC$(3,6,L,3)$ ensemble via suitable channel interleaving. The region is plotted for $L=25,50$
and $100$. }
\end{figure}

One can show that the DE equation is monotonically increasing in terms of $\e_z$.
Therefore, if $(\e_1,\e_2)$ is achievable under BP decoding, then for all $\e'_1\leq \e_1$ and $\e'_2\leq\e_2$, $(\e'_1,\e'_2)$ is also
achievable under BP decoding (with different $(\mu_1,\mu_2,B)$).  
Using \eqref{eq:g_(B,mu)}, we can numerically calculate the maximal achievable region of $(\e_1,\e_2)$ which can be decoded successfully in the limit of the code-length.
We plot this achievable region in Fig.~\ref{fig:areas} for the tail-biting SC-LDPC$(3,6,L,3)$ ensemble with $L=25$, $50$ and $100$. 
The  achievable region under MAP decoding is $\e_1+\e_2\leq 2\e_\MAP$. We observe that
by suitable interleaving, the achievable region under BP decoding almost covers the entire MAP achievable region except for a small triangular region in which $\e_1\geq\e_\BP$ and $\e_2\geq\e_\BP$. As $L$ grows, the coverage
saturates to the MAP achievable region (except for  $\e_1\geq\e_\BP$ and $\e_2\geq\e_\BP$). 

\subsection{4-ASK BICM over the AWGN Channel}

Bit-interleaved coded-modulation (BICM) scheme is a 
practical pragmatic scheme to approach the \emph{constellation constrained capacity} of the AWGN channels with higher order modulation formats~\cite{caire1998bit}. In BICM over a real-valued channel, we map $k$ (possibly interleaved) code bits to $2^k$ symbols of an Amplitude Shift-Keying (ASK) constellation $\{\pm 1, \pm 3, \ldots, \pm (2^{k-1}+1)\}$ using a Gray map.
The LLRs of the transmitted $k$ code bits are computed using a so-called \emph{bit-metric decoder}: The LLRs are generally different for each code bit and some of the bits may be more reliable than others, depending on the mapping used. Interleaving is used to break the correlation between the different bits.

For a tail-biting SC-LDPC code, this interleaving  can be carried out separately in each spatial position. Assuming uncorrelated LLRs,
we will have $k$ parallel channels with different LLR distributions~\cite{caire1998bit}. It has been shown in \cite{yedla2013performance,schmalen2013combining} that we can approach 
the BICM capacity using only a single (terminated) SC-LDPC code. 
A suitable interleaving scheme can significantly improve the BP performance of tail-biting SC-LDPC codes over
such parallel channels with different reliability as occurring in BICM. More interestingly, such a code can outperform the terminated
SC-LDPC code with the same design parameters and with \emph{the same energy per transmitted bits}, as we will show below.
Note that generally speaking, we need to consider two distinct interleaving schemes, which are however usually combined in practice: One is required to remove the correlation for BICM, and the other one, which we consider, enables threshold saturation. 

Here we focus on the $4-$ASK modulation scheme\footnote{4-ASK modulation represents one quadrature component in a 16-QAM modulation, the results discussed here therefore directly apply to 16-QAM.} over an AWGN 
channel with SNR $\eta=E_s/N_0$. BICM yields two binary 
parallel channels, namely $c_1(\eta)$ and $c_2(\eta)$, with different LLR distributions. To use DE, the channels must be symmetric, which can be ensured by ``channel adaptation'' as detailed in \cite{hou2003capacity}. 

Recall that $\beta_{z}$ is the fraction of code bits in spatial position $z$ 
transmitted over $c_1(\eta)$ (with constraint $\sum_{z=0}^{L-1}\beta_z=L/2$). Let $P_e(\ub,T)$ denote the error probability of BP decoding after $T$ iterations. We define
\begin{equation}\label{eq:beta_opt}
\ub_{\rm opt}=\arg\min_{\ub}\{\eta\mid \lim_{T\to\infty} P_{\rm e}(\ub,T)=0\}.
\end{equation}
Similar as in Section~\ref{sec:random_shortening}, 
we relax the optimization problem by introducing a new condition $ P_{\rm e}(\ub,T)<10^{-8}$
for $T<1000$ iterations. We consider two distinct setups, first  uniform interleaving with
\begin{equation*}
\ub_{\rm uni}(B) =
\begin{cases}
\beta_0,&0\le z< B\\
\frac{\frac{L}{2}-B\beta_0}{L-B},& z\ge B
\end{cases}
\end{equation*}
and second non-uniform interleaving with $\ub=(\beta_0,\beta_1,...,\beta_{L-1})$ and find the minimum $\eta$ for both setups. For  uniform interleaving, we select the minimum $\eta$ over all $0<B<L$.
For non-uniform interleaving, we use the technique of differential evolution to find the best sub-optimal $\ub^*$, similarly to~\cite{hager2014optimized}. To evaluate $P_{\rm e}(\ub,T)$, we numerically compute the LLR distributions 
of $c_1(\eta)$ and $c_2(\eta)$ for each $\eta$ and then use quantized DE analysis (uniformly quantizing the LLR range $[-20,20]$ with $1000$ levels). This method is more accurate than  the P-EXIT method and Gaussian approximation of channel's LLR distributions
used in \cite{hager2014optimized,hager2015terminated} and may potentially lead to improved results, however, with a significantly increased (offline optimization) complexity.

As an example, consider the tail-biting SC-LDPC(3,6,50,3) ensemble. Without interleaving ($\beta_z=\frac{1}{2}$), the BP threshold is
$\eta=3.399$\,dB while the BP threshold of the terminated ensemble is
$\eta=2.658$\,dB. By a uniform interleaving, the BP threshold 
of tail-biting ensemble improves to $\eta=2.809$\,dB and it is even slightly more improved to $\eta=2.804$\,dB by the best non-uniform interleaving we have found.

\begin{figure}[tb]
\centering
\includegraphics[width=\columnwidth]{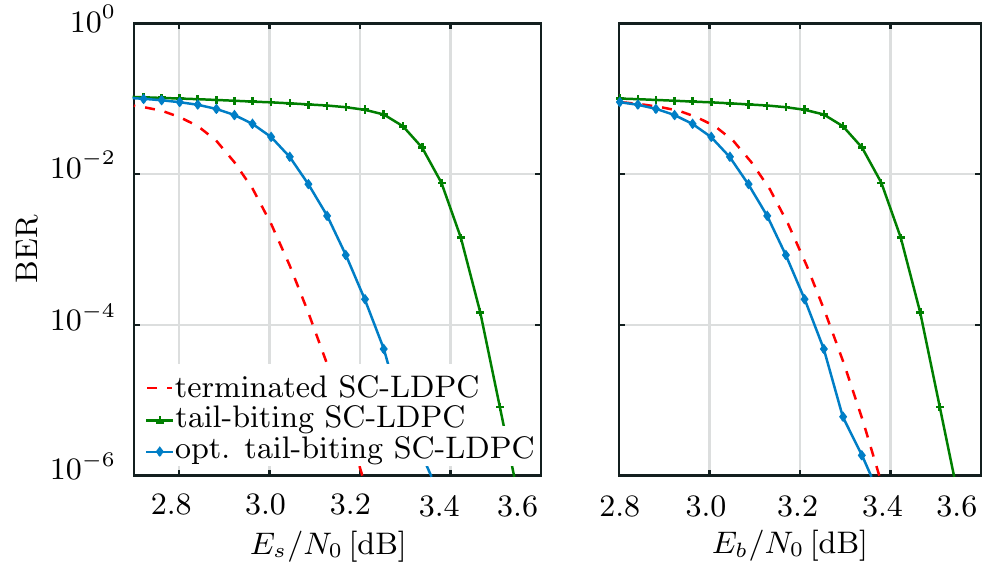}
\caption{BP performance of the 4-ASK BICM scheme over the AWGN channel using an optimized tail-biting SC-LDPC code (blue curve), a terminated SC-LDPC code (red curve), and a tail-biting SC-LDPC code (green curve). For all codes, 
$(d_v,d_c,L,w,n)=(3,6,50,3,2000)$. 
Left: BER values in terms of $E_s/N_0$. Right:
BER values in terms of $E_b/N_0.$}
\label{fig:results_2ch_ber_awgn}
\end{figure}

Since the terminated ensemble has a smaller rate, it is more fair to compare
the BP thresholds in terms of $E_b/N_0$. The rate of tail-biting ensemble is $1/2$ and 
thus, $E_b/N_0=E_s/N_0$. However, the rate of the terminated ensemble is $0.481$
and thus its BP threshold is $E_b/N_0=2.826$\,dB, larger than the BP threshold of tail-biting ensemble with optimal interleaving. This 
superior performance can be also seen in Fig.~\ref{fig:results_2ch_ber_awgn} comparing the averaged BP performance of actual codes with finite $n=2000$.
In terms of $E_b/N_0$, we observe a net coding gain of $\approx 0.04$\,dB in this simulation.

\section{Encoding Queue For Bit Interleaving}
The convolutional structure of SC-LDPC codes (including tail-biting codes)
enables sequential encoding of information bits as they are produced with a structure akin to shift registers. In particular, the code bits in spatial position $z$ are generated from 
the code bits in position $[z-w,z]$.\footnote{The range of required information bits may change for different encoders. It is important that its size is $O(w)\ll L$.} On the other hand, the windowed BP decoder
enables sequential decoding of code bits as they arrive at the receiver.
These features make SC-LDPC codes particularly attractive for high-speed, latency sensitive applications~\cite{schmalen2015spatially}.

If we assume that the parallel channels should be used an equal amount of time, i.e., the gross transmission rate of the channels is identical, code-bit interleaving between spatial positions violates the equal 
fraction of code bits targeted for different channels. For example, in BICM, we always require an equal amount of all channels to be used in every frame to generate the modulation symbols. Therefore, some code bit must be stored in a buffer and transmitted later. This drawback causes queuing of code bits in the encoder and decoder and ultimately leads to an increase in transmission latency.\footnote{Note that we do not include the additional buffering required for tail-biting in our considerations here, as this amount stays constant irrespective of the interleaving.}  

In this section, we consider the size of this encoding queue as an additional constraint when optimizing the channel interleaving.
For a given pair of channels, there are usually many $\ub$ leading to practically the same ``quasi-optimal'' BP performance. Our goal is to find the one that leads to the encoding queue of minimum size.

For simplicity, we focus here on two parallel BECs but the results can be carried immediately over to other channel models, such as BICM as shown above. 
We assume that the code bits are generated sequentially from position $z=0$ to $z=L-1$ with a constant speed. We assume that the encoder processed each spatial position as a block. Recalling that $\beta_z$ is the fraction of code bits in position $z$ to be transmitted over the first channel, then it is clear that only $n\cdot\min(\beta_z,1-\beta_z)$ bits can be forwarded to the parallel channels to be mapped onto modulation symbols and the remaining $n|1-2\beta_z|$ bits need to be stored in the queue, waiting further processing. Let $Q(z)= n\sum_{i=0}^z(2\beta_i-1)$ denote the length (in bits) of the encoding queue after encoding spatial positions up to $z$, where $Q(z) > 0$ denotes an excess of first channel bits and $Q(z) < 0$ an excess of second channel bits. The total required memory size for the queue is $Q(\ub)=\max_z\vert Q(z)\vert$
which also shows the extra encoding latency. 

\begin{figure}[tb]
\centering
\includegraphics[width=\columnwidth]{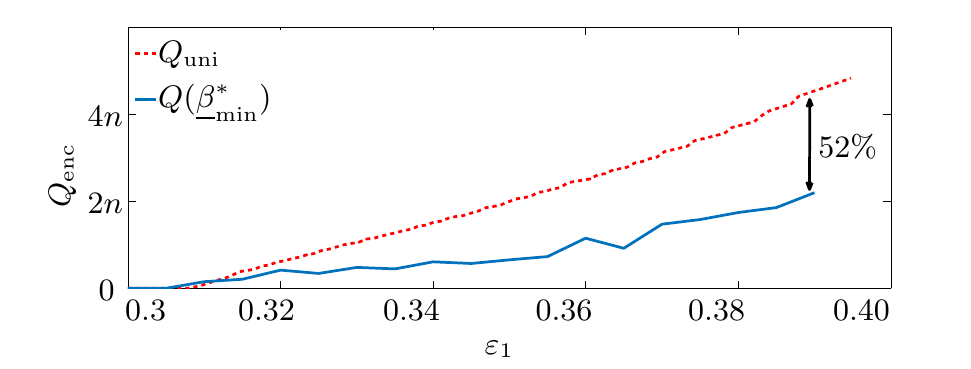}
\caption{\label{fig:enc_que_2d} The encoding queue length $Q(\ub)$ of a tail-biting SC-LDPC(3,6,50,3) ensemble over BECs $(\e_1,\e_2=0.55)$ and $\e_1\in(0.3,0.39)$.}
\end{figure}

\begin{figure}[tb]
\includegraphics[width=\columnwidth]{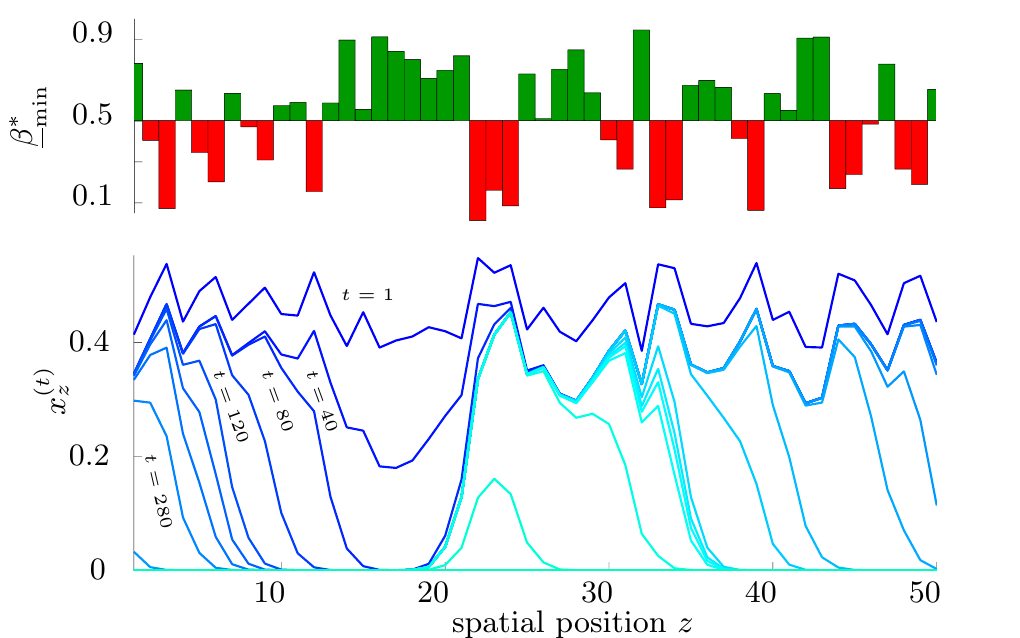}
\caption{\label{fig:wave_enc_que} The evolution of densities $x_z^{(t)}$ for the tail-biting SC-LDPC$(3,6,50,3)$ ensemble over two BECs $(\e_1=0.375, \e_2=0.55)$ and with the optimized non-uniform interleaving $\ub_{\rm min}^*$. Each curve represents $x_z^{(t)}$ after $t=1,40,80,...$ iterations.}
\end{figure}

Consider now a pair of BECs $(\e_1,\e_2)$ in the achievable region of Fig.~\ref{fig:areas}. Our objective is to minimize the queue, i.e.,
\begin{equation}
\ub_{\rm min}=\arg\min_{\ub}\{Q(\ub)\vert \lim_{T\to\infty}P_e(\ub,T)=0\}.
\end{equation}
We relax the above optimization similarly as in~\eqref{eq:relaxed_Pe} and we consider again both the uniform interleaving
$\ub_{\rm min}(B)$,
and the non-uniform interleaving in which we use the differential evolution
to find $\ub_{\rm min}^*$ with the smallest queue. For uniform interleaving,
we choose the minimum queue $\ub_{\rm min,uni}(B)$ over all $B$.
The optimization results for the tail-biting SC-LDPC(3,6,50,3) ensemble are depicted in Fig.~\ref{fig:enc_que_2d}. For the sake of a clear presentation, we fix $\e_2=0.55$ and
find the minimum queue size for $0.3<\e_1<0.39$. As we can see, 
the required queuing memory can be significantly smaller using a non-uniform interleaving. 
As discussed in Section~\ref{sec:two_becs}, the optimal non-uniform
interleaving improves the BP performance only slightly in comparison to the 
optimal uniform interleaving. However, it can have many practical advantages such as large memory savings for realizing the encoding queue. Fig.~\ref{fig:wave_enc_que} shows the non-uniform interleaving found for $(\e_1=0.375, \e_2=0.55)$. It is interesting that as a result of such interleaving,
the average erasure probability $\e_z=\beta_z\e_1+(1-\beta_z)\e_2>\e_{\MAP}$ in some spatial positions.

\section{Conclusion}
\label{sec:conclusion}

Rate-loss mitigation of spatially coupled codes is one of the major challenges towards a practical implementation of this class of codes.
We have shown that a significant rate-loss reduction
can be obtained from shortening tail-biting SC-LDPC codes. For the case of SC-LDPC$(d_v=3,d_c=6,L,w=3)$ codes,
we can reduce the rate-loss by more than $50\%$ by a suitable shortening pattern. The shortened tail-biting SC-LDPC codes can outperform terminated SC-LDPC codes for transmission over AWGN channels.

Rate-loss mitigation can be fulfilled effectively when the transmission takes place over more than  
one binary-input channel, as is the case in BICM. The extra available channel dimensions can be exploited by properly
interleaving code bits in different spatial positions among channels.
For the case of two parallel BECs, we show that a tail-biting SC-LDPC code under BP decoding can be operated almost anywhere within the achievable region of MAP decoding.
Optimized bit mappings for higher order modulation can help to increase the spectral efficiency of the code. However, the price is a mandatory buffer at the encoder. This effect can be mitigated by non-uniform channel allocation patterns, which show a better performance than the uniform schemes.

\bibliographystyle{IEEEtran}

\newcommand{\SortNoop}[1]{}

\end{document}